\documentclass[preprint, 3p]{elsarticle}

\usepackage[T1]{fontenc} 

\usepackage{siunitx} 			
\usepackage{hepnames}         	
\usepackage{hepparticles}       
\usepackage{graphicx}           
\usepackage{booktabs}           
\usepackage{pdflscape}          
\usepackage{subcaption}         
\usepackage{aas_macros}         
\usepackage{mathrsfs}           
\usepackage{lineno}             
\usepackage{placeins}           
\usepackage{nicefrac}           

\newcommand{\jann}{\ensuremath{J_{\text{ann}}}} 

\begin{document}

\begin{frontmatter}

\title{A search for dark matter in Triangulum~II with the MAGIC telescopes}

\author[a]{\textbf{MAGIC Collaboration}: V.~A.~Acciari}
\author[b,w]{S.~Ansoldi}
\author[c]{L.~A.~Antonelli}
\author[d]{A.~Arbet Engels}
\author[e]{D.~Baack}
\author[f]{A.~Babi\'c}
\author[g]{B.~Banerjee}
\author[h]{U.~Barres de Almeida}
\author[i]{J.~A.~Barrio}
\author[a]{J.~Becerra Gonz\'alez}
% ...

\address[a]{Inst. de Astrof\'isica de Canarias, E-38200 La Laguna, and Universidad de La Laguna, Dpto. Astrof\'isica, E-38206 La Laguna, Tenerife, Spain}
\address[b]{Universit\`a di Udine, and INFN Trieste, I-33100 Udine, Italy}
\address[c]{National Institute for Astrophysics (INAF), I-00136 Rome, Italy}
\address[d]{ETH Zurich, CH-8093 Zurich, Switzerland}
\address[e]{Technische Universit\"at Dortmund, D-44221 Dortmund, Germany}
\address[f]{Croatian Consortium: University of Rijeka, Department of Physics, 51000 Rijeka; University of Split - FESB, 21000 Split; University of Zagreb - FER, 10000 Zagreb; University of Osijek, 31000 Osijek; Rudjer Boskovic Institute, 10000 Zagreb, Croatia}
\address[g]{Saha Institute of Nuclear Physics, HBNI, 1/AF Bidhannagar, Salt Lake, Sector-1, Kolkata 700064, India}
\address[h]{Centro Brasileiro de Pesquisas F\'isicas (CBPF), 22290-180 URCA, Rio de Janeiro (RJ), Brasil}
\address[i]{IPARCOS Institute and EMFTEL Department, Universidad Complutense de Madrid, E-28040 Madrid, Spain}
% ...
\address[w]{Japanese MAGIC Consortium: ICRR, The University of Tokyo, 277-8582 Chiba, Japan; Department of Physics, Kyoto University, 606-8502 Kyoto, Japan; Tokai University, 259-1292 Kanagawa, Japan; RIKEN, 351-0198 Saitama, Japan}
% ...

\cortext[cor1]{Corresponding authors: jarred.green@inaf.it, francesco.saturni@inaf.it, jpalacio@ifae.es, michele.doro@unipd.it}

\begin{abstract}
We present the first results from very-high-energy observations of the dwarf spheroidal satellite candidate Triangulum~II with the MAGIC telescopes from 62.4 hours of good-quality data taken between August 2016 and August 2017. We find no gamma-ray excess in the direction of Triangulum~II, and upper limits on both the differential and integral gamma-ray flux are presented. Currently, the kinematics of Triangulum~II are affected by large uncertainties leading to a bias in the determination of the properties of its dark matter halo. Using a scaling relation between the annihilation \emph{J}-factor and heliocentric distance of well-known dwarf spheroidal galaxies, we estimate an annihilation \emph{J}-factor for Triangulum~II for WIMP dark matter of $\log[\jann({0.5^{\circ}})/$~GeV$^{2}$~cm$^{-5}] = 19.35 \pm 0.37$. We also derive a dark matter density profile for the object relying on results from resolved simulations of Milky Way sized dark matter halos. We obtain 95\% confidence-level limits on the thermally averaged annihilation cross section for WIMP annihilation into various Standard Model channels. The most stringent limits are obtained in the \HepProcess{\Ptauon\APtauon} final state, where a cross section for annihilation down to $\langle \sigma_{\text{ann}} v \rangle = 3.05 \times 10^{-24}$~cm$^{3}$~s$^{-1}$ is excluded.
\end{abstract}

\begin{keyword}
dark matter \sep indirect searches \sep dwarf spheroidal satellite galaxies \sep Imaging Air Cherenkov Telescopes \sep Triangulum II
\end{keyword}

\end{frontmatter}


\section{Introduction} 
\label{sec:introduction}

The dark matter (DM) paradigm arises from observational evidence 
which shows the Standard Model (SM) of particle physics cannot entirely explain the gravitational effects on astrophysical systems observed at all cosmological scales, from Milky Way (MW) satellite dwarf spheroidal galaxies (dSphs) to clusters of galaxies~\cite{2010arXiv1001.0316R,2009EAS....36..113F}. DM is expected to represent nearly $85\%$ of the total matter content of our universe, however its true nature remains elusive. The existence of one or more new massive particles not belonging to the SM is a well-favored explanation. Experimental evidence constrains the DM particle to be stable on cosmological timescales, electrically neutral, and non-baryonic. Weakly interacting massive particles (WIMPs) fulfill all of these requirements and are among the best-motivated DM candidates~\citep{1977PhLB...69...85H,2012PhRvD..86b3506S}. Expected to have masses in the range of a few \si{GeV} to a few \si{TeV}, WIMPs interact with SM particles at most on the weak scale, and could explain the observed relic density of ${\Omega_{DM}=0.259\pm 0.006}$~\citep{2018arXiv180706209P}. WIMP particles can interact and produce various SM particles, possibly including gamma rays that could be detected by ground- and space-based observatories. In spite of the various efforts that have been performed, no evidence for the existence of DM particles has been found.

The Florian Goebel Major Atmospheric Gamma-ray Imaging Cherenkov (MAGIC) telescope is a pair of Imaging Atmospheric Cherenkov Telescopes (IACTs), each 17 m in diameter, located at the Roque de los Muchachos Observatory (Spain). The system is sensitive in the very-high-energy (VHE) gamma-ray regime. MAGIC is able to detect gamma rays via the Cherenkov light produced during atmospheric showers initiated when VHE photons interact with the Earth's atmosphere. For low zenith angle observations, MAGIC has standard trigger threshold of $\sim50$~\si{GeV}, a 68\% containment angular point spread function (PSF) of~$\sim0.1$ deg for gamma rays of $\sim200$~\si{GeV}, and an energy resolution of 16\%~\cite{2016APh....72...61A}, rendering the instrument particularly well-suited to perform indirect DM searches. Among the most promising astrophysical objects for indirect DM searches with IACTs are dSphs~\cite{2014NIMPA.742...99D,2016JCAP...07..025U}. These compact galaxies are gravitationally bound to the MW and located in its galactic halo. They have very low luminosities, few member stars, no gas or dust, and tend to be approximately spheroidal in shape. In addition, a low astrophysical background makes them optimal targets for searches in the gamma-ray regime~\cite{2008ApJ...678..614S}. They have among the highest mass-to-light ratios (M/L) of any known astrophysical object, and many are found relatively nearby (within 250~\si{kpc} of the center of the MW).

The first VHE indirect DM search based on the observation campaign of Triangulum~II (Tri~II), a relatively nearby and a potential DM-dense dSph~\citep{2015ApJ...814L...7K,2017ApJ...838...83K}, was carried out with the MAGIC Telescopes and is reported in this paper. These observations are part of a multi-year MAGIC campaign of dSphs~\cite{2014JCAP...02..008A,2018JCAP...03..009A}, which follows a common analysis likelihood framework~\cite{2016JCAP...02..039M}. The main difference with respect to previous MAGIC dSph studies lies in the determination of the \emph{J}-factor and emission morphology of Tri~II, due to an absence of sufficient stellar kinematic data.


\section{Standard gamma-ray analysis of Triangulum~II} 
\label{sec:the_observation_campaign_of_triangum_ii_with_magic}

Triangulum~II is an ultra-faint MW satellite discovered in 2015 as part of the Pan-STARRS survey~\citep{2010SPIE.7733E..0EK}. It is located at RA(J2000) 02h13m17.4s, Dec(J2000) +36$^{\circ}$10'42.4'', with an absolute magnitude of ${M_{v} = -1.8 \pm 0.5}$ at a heliocentric distance of $d = {30 \pm 2}$~\si{kpc}~\citep{2015ApJ...802L..18L}. A first spectroscopic study, based on six member stars, predicted Tri~II to have a M/L of ${3600^{+3500}_{-2100} \ M_{\odot}/L_{\odot}}$ and one of the highest DM concentrations of any galaxy ever found~\citep{2015ApJ...814L...7K}.

An observation campaign of Tri~II was subsequently carried out with the MAGIC telescope. MAGIC collected 62.4~\si{h} of good-quality data between August 2016 and August 2017 as part of a multi-year campaign to study different dSphs~\citep{2014JCAP...02..008A,2018JCAP...03..009A}.  The Tri~II data were collected at low zenith angles, between ${{\sim}\ang{5}}$ and ${{\sim}\ang{35}}$. The observations were carried out in `wobble' mode, where the target was tracked with a ${0.4^{\circ}}$ offset from the center of the camera~\citep{1994APh.....2..137F}, allowing for the simultaneous measurement of both the source (`ON') and the background control (`OFF') regions. Two telescope pointings were used, lying on an axis inclined ${148^{\circ}}$ with respect to the line of constant declination at the Tri~II position (which prevented the relatively bright nearby star $\beta$ Trianguli, $m=3.02$, from illuminating the trigger area of the MAGIC cameras~\citep[][]{2016APh....72...61A}).

More recent kinematic studies of Tri~II, carried out after the MAGIC observation campaign, indicate that Tri~II might not be in dynamical equilibrium~\citep{2017ApJ...838...83K}. Further spectral measurements with Keck/DEIMOS of 13 member stars hint at the fact that the object is currently in the process of disruption and is not virialized as a result, or possibly embedded in a stellar stream~\citep{2016ApJ...818...40M,2003SPIE.4841.1657F}. Further studies suggest that one of the original stars used in kinematic calculations was actually a binary system, which artificially increased the total velocity dispersion thus wrongfully boosting the estimated DM density in~\citep{2017ApJ...838...83K}.

\begin{figure}[!tb]
	\centering
	\includegraphics[width=0.75\textwidth]{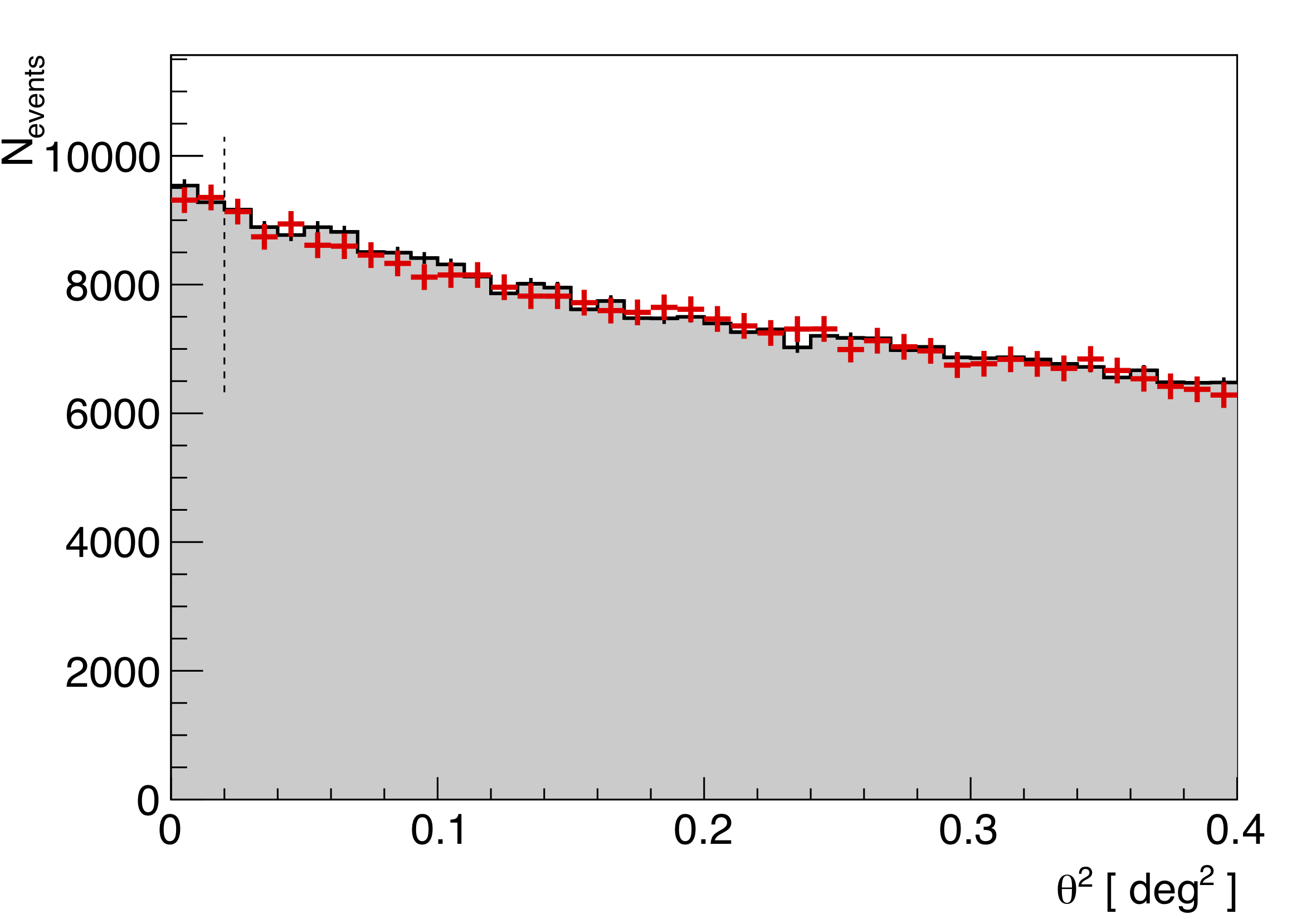}
	\caption{The $\theta^{2}$ distribution of gamma-like events in the ON (red crosses) and OFF (shaded) regions for Tri~II. The vertical dashed line at $\theta^{2} = 0.02$~\si{deg^{2}} shows the border of the fiducial signal region defined for a point-like source. The distribution of events in the OFF region has been scaled to match that of the ON region at distances far outside this fiducial signal region. There is no significant detection of gamma-ray signal over the background.}
	\label{fig:theta2}
\end{figure}

\begin{figure}[!tb]
	\centering
	\includegraphics[width=0.65\textwidth]{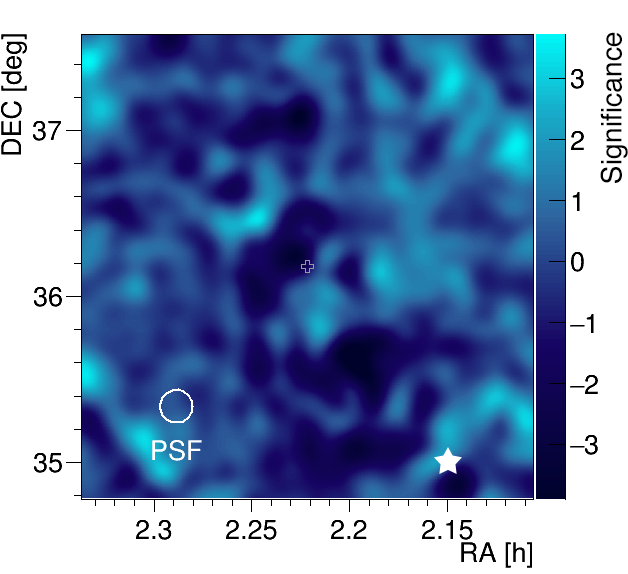}
	\caption{Significance sky-map centered at the Tri~II sky position (white cross). No significant gamma-ray signal is detected over the background. The color scale represents the test significance. The white circle shows the average MAGIC PSF (68\% containment) for the event selection used in the analysis (see Section~\ref{sub:flux_upper_limits}). The location of the star $\beta$ Trianguli is shown as a white star.}
	\label{fig:skymap}
\end{figure}

The data calibration and analysis was carried out using the standard MAGIC Analysis and Reconstruction Software (MARS)~\citep{mars_zanin}. The total sample of data underwent a selection process for excellent atmospheric conditions, quantified by a proprietary LIDAR sensing instrument~\citep{2014arXiv1403.3591F}. Events surviving the aforementioned data selection criteria are assigned an estimated energy and direction, as well as a gamma/hadron discriminator called ``hadronness'' or $h$. $h$ is a computed by the comparison of real data with dedicated Monte Carlo (MC) gamma-ray simulations using the Random Forest method~\cite{2008NIMPA.588..424A}.

In Figure~\ref{fig:theta2}, following the standard low-energy analysis of MAGIC, we show the distribution of the number of gamma-like events as a function of $\theta^{2}$, the squared angular distance between the reconstructed event direction and the nominal position of the target, around the ON and OFF regions\footnote{\label{note:off_region}Distribution of events in the OFF region has been scaled to match the one from the ON region at distances much larger than our fiducial signal region in $\theta^{2}$.}. No excess gamma-ray signal is detected over the background in the direction of Tri~II. 

A significance map centered in the target sky position of Tri~II is shown in Figure~\ref{fig:skymap}. This sky-map is generated using a test statistic according to~\cite{1983ApJ...272..317L} (Eq.~17) and is applied on a smoothed, modeled background estimation~\citep{2011ICRC....3..266L}. Again, no hint of emission is seen from the region of Tri~II.

\subsection{Gamma-ray flux limits} 
\label{sub:flux_upper_limits}
\begin{figure}[!tb]
	\centering
	\includegraphics[width=\textwidth]{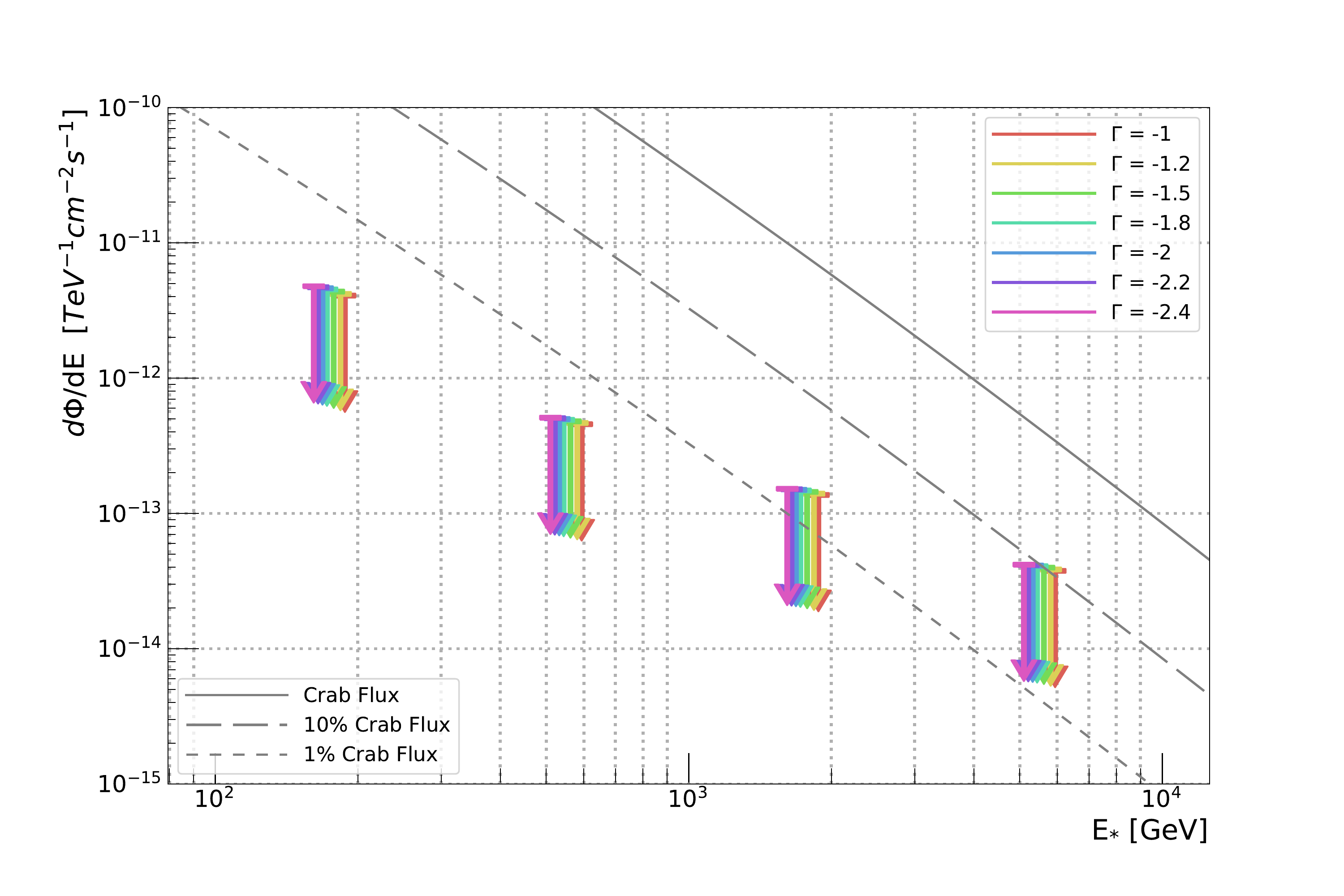}
	\caption{Differential gamma-ray flux upper limits of Tri~II, assuming a point-like source and a power-law emission spectrum with various spectral slopes $\Gamma$. The solid gray line represents the Crab Nebula flux as detected by MAGIC, as well as the 10\% and 1\% of the Crab Nebula flux for reference. The x-axis is defined in terms of the pivot energy $E^{*}$, calculated for each energy bin (see Equation A.2 of ~\cite{2014JCAP...02..008A}). The numerical values of the obtained flux ULs are shown in~\ref{sec:gamma_ray_flux_upper_limits}.} 
	\label{fig:Triangulum_Differential_Upper_Limits}
\end{figure}

We derive 95\% confidence level (CL) upper limits (UL) on the gamma-ray flux from the region of Tri~II. We follow the same framework as adopted from the MAGIC Segue~I dSph analysis~\citep{2014JCAP...02..008A}, that made use of power law spectra with 7 different values of the spectral index $\Gamma$ ranging between $-1$ and $-2.4$. We select events based on h retaining 80\% MC gamma rays in each energy bin and define a fiducial signal region until $\theta^{2} = 0.02$~\si{deg^{2}} (these cuts are used through the rest of this paper unless otherwise specified). The differential flux ULs in the energy range 100~\si{GeV} to 10~\si{TeV} are shown for each assumed spectral index in Figure~\ref{fig:Triangulum_Differential_Upper_Limits}, and the obtained ULs are compared, for reference, to the Crab Nebula flux as measured by MAGIC~\citep{2008ApJ...674.1037A}. The x-axis is defined in terms of the pivot energy $E^{*}$, which is calculated for each energy bin (see Equation A.2 of~\cite{2014JCAP...02..008A}).

\begin{figure}[!tb]
	\centering
	\includegraphics[width=\textwidth]{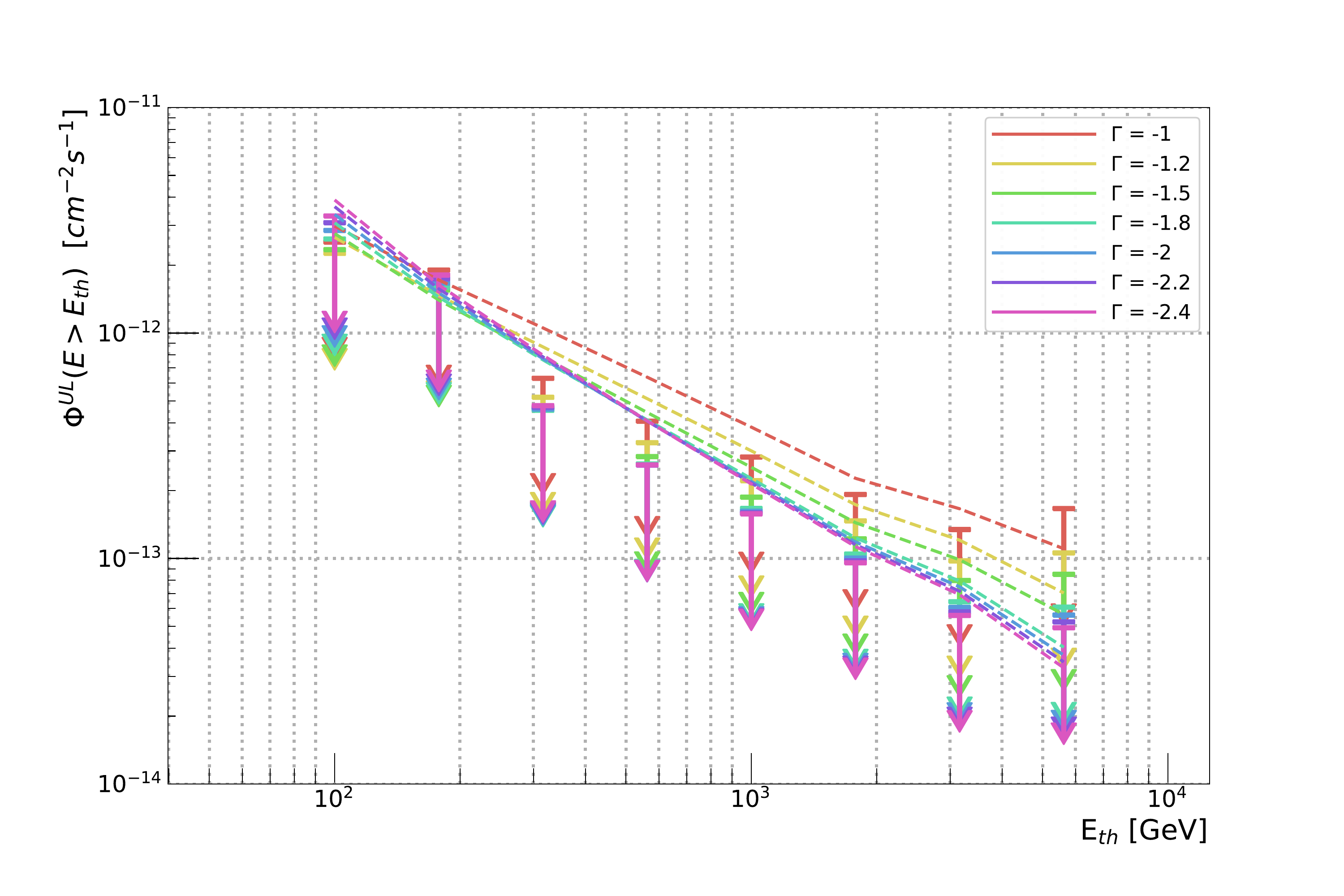}
	\caption{Integral gamma-ray flux upper limits of Tri~II for events with energy $E > E_{\text{th}}$, assuming a point-like source and a power-law emission spectrum with various spectral slopes $\Gamma$. The `null hypothesis' case is represented by dashed lines and the numerical values of the obtained flux ULs are shown in~\ref{sec:gamma_ray_flux_upper_limits}.}
	\label{fig:Triangulum_Integral_Upper_Limits}
\end{figure}	

The integral flux ULs for gamma rays with $E > E_{\text{th}}$ are shown in Figure~\ref{fig:Triangulum_Integral_Upper_Limits}, for various assumed energy thresholds $E_{\text{th}}$. We note that, as expected, the integral ULs are more dependent on the assumed spectral index and are more sensitive to statistical fluctuations than the differential limits. To quantify this effect, we also calculate the expected ULs under a `null-hypothesis' ${N^{\text{UL}}_{\text{ex}}(E>E_{\text{th}}) = 0}$, shown as dashed lines in Figure~\ref{fig:Triangulum_Integral_Upper_Limits}. In this case, a significance of zero (i.e. when the signal region of the $\theta^{2}$ plot contains a number of events in the ON region equal to the scaled number of events in the OFF region) is assumed for each value of $\Gamma$ and $E_{\text{th}}$. The numerical values of the obtained flux ULs of Figure~\ref{fig:Triangulum_Differential_Upper_Limits} and Figure~\ref{fig:Triangulum_Integral_Upper_Limits} are shown in~\ref{sec:gamma_ray_flux_upper_limits}.


\section{Dark matter searches in Triangulum~II} 
\label{sec:dark_matter_search}

We perform a dedicated analysis of gamma-ray signatures of DM particles annihilating into SM pairs, taking advantage of the MAGIC observations of Tri~II presented in Section~\ref{sec:the_observation_campaign_of_triangum_ii_with_magic}. The detection of a potential signal in the gamma-ray energy window $100~\si{GeV}$ -- $100~\si{TeV}$ would provide information both on the nature of the DM progenitor particles and on the DM distribution in the direction of the source. In the case that no signal is detected, constraints on the annihilation cross section of DM can be obtained, allowing for an exploration of the parameter space of potential candidate particles. 

The expected gamma-ray flux from DM annihilation in nearby sources can be expressed as the product of two terms:
\begin{equation}
	\frac{d\Phi_{\gamma}}{dE} = \frac{d\Phi_{\gamma}^{PP}}{dE} J_{\text{ann}}(\Omega) \,\,,
\end{equation}
where $\nicefrac{d\Phi_{\gamma}^{PP}}{dE}$ and $J_{\text{ann}}(\Omega)$ are usually referred to as the particle physics factor and the astrophysical factor, respectively. The factor $\nicefrac{d\Phi_{\gamma}^{PP}}{dE}$ contains all information related to the particle behavior:
\begin{equation}\label{eq:annihilation_PP_factor}
	\frac{d\Phi_{\gamma}^{PP}}{dE} = \frac{\langle \sigma_{\text{ann}} v \rangle}{8\pi m^{2}_{DM}} \frac{dN_{\gamma}}{dE} \,\,,
\end{equation}
where $\langle \sigma_{\text{ann}} v \rangle$ is the thermally averaged DM annihilation cross section, $m_{DM}$ is the DM particle mass, and $\nicefrac{dN_{\gamma}}{dE}$ provides the average spectrum of photons expected per each annihilation interaction. The astrophysical factor $J_{\text{ann}}(\Omega)$, also known as the \emph{J}-factor, contains information on how DM is distributed in the source with respect to earth~\citep{1998APh.....9..137B}, and is calculated via the line-of-sight (l.o.s.) integral of the square of the DM density distribution:
\begin{equation}\label{eq:jfactor_definition}
	J_{\text{ann}}(\Delta\Omega) = \int_{\Delta\Omega} \int_{los} \rho^{2}(r(l))\, dl\, d\Omega \,\,,
\end{equation}
where $l$, $\rho(r(l))$ and $\Delta\Omega$ are the l.o.s. element, density profile of DM within the source, and the solid angle of the integrated region, respectively.

\subsection{Determination of $J_{\text{ann}}$} 
\label{sub:determination_of_jann}

Due to the lack of sufficient spectroscopic data for Tri~II, a precise determination of $\rho(r)$, and hence \jann\ (through Eq.~\ref{eq:jfactor_definition}), cannot be made. Therefore, we estimate \jann\ with an alternative approach based on scaling relations among various observables inferred from other dSphs~\citep{2017ApJ...834..110A,2019MNRAS.482.3480P}.

Several studies suggest the existence of a number of scaling relations between various observable characteristics of dSphs and their DM content. It is observed that many dSphs tend to have similar integrated DM masses independent of their luminosities~\citep{2008Natur.454.1096S,2009ApJ...704.1274W,2010MNRAS.406.1220W}. This leads to a simple relation that links the heliocentric distance of the dSph $d$ to its value of \jann, scaling as ${\jann\ \propto d^{-2}}$~\citep{2015ApJ...809L...4D,2017ApJ...834..110A}. Note that estimates on $d$ are generally derived with variable stars or isochrone fitting and do not rely on stellar kinematic information. Using statistical models which involve the fitting of parameters of a large number of dSphs, Pace \& Strigari~\cite{2019MNRAS.482.3480P} find that the best fit of the ${\jann\ \propto d^{-2}}$ scaling relation is:
\begin{equation}\label{eq:scaling_relation}
	\frac{\jann(0.5^{\circ})}{\si{GeV^2}\si{cm}^{-5}} = 10^{18.30 \pm 0.07} \left(\frac{d}{100\si{kpc}}\right)^{-2} \, ,
\end{equation}
with an intrinsic scatter of $\sigma_{\log_{10}J_{\text{ann}}} = 0.30 \pm 0.07$ and where \jann\ is integrated out to $0.5^{\circ}$ from the center of each dSph. Here, the intrinsic scatter quantifies the spread of the parent population around an average relation due to poorly constrained or unknown dependencies on physical parameters that characterize the phase space of the sample. We also note that the reasons for selecting a \emph{J}-factor scaling relation which includes DM halos out to $0.5^{\circ}$ from the center of each dSph is twofold: first, Pace \& Strigari~\cite{2019MNRAS.482.3480P} find that at this angle the scaling relation has the lowest scatter compared to a variety of other integration angles. Second, the majority of the DM content of the dSphs used to calculate the relation is contained within an integral angle of $0.5^{\circ}$ (e.g. see Table A2 of~\cite{2019MNRAS.482.3480P} and Table 2 of~\cite{2015MNRAS.453..849B}).

\begin{sloppypar}
Based on its distance of $d = 30 \pm 2$~\si{kpc}, Tri~II is calculated to have a \emph{J}-factor of ${\log_{10}[\jann(0.5^{\circ}) / \si{GeV^2}\si{cm}^{-5}] = 19.35 \pm 0.37}$. The uncertainty on \jann\ is computed by adding the $1\sigma$ extent of the intrinsic scatter to the error of Eq.~\ref{eq:scaling_relation}, and must be treated carefully. We consider this uncertainty to be reasonable under the assumption that the physical properties of Tri~II align with the dSph scaling relations in the literature. Note that this value for \jann\ is also consistent with other scaling relations, such as the luminosity scaling relation of ~\cite{2019MNRAS.482.3480P}.
\end{sloppypar}

In general, the DM distribution of a source can be reconstructed from good-quality spectroscopic information. Since this is not possible for Tri~II based on predictions by numerical N-body DM simulations (as discussed in Section~\ref{sec:the_observation_campaign_of_triangum_ii_with_magic}), we choose to estimate the DM density profile of Tri~II.

We first adopt an Einasto profile~\citep{1965TrAlm...5...87E} for the DM density of Tri~II:
\begin{equation}\label{eq:einasto}
	\rho(r)=\rho_{-2}\exp\left\{-\frac{2}{\alpha}\left[\left(\frac{r}{r_{-2}}\right)^{\alpha}-1\right]\right\} .
\end{equation}
Here, the parameters $\rho_{-2}$ and $r_{-2}$ represent the density and radius at which the local slope of $\rho_{r}$ is equal to $-2$, respectively. We fix the profile index $\alpha = 0.16$. Such a choice is motivated by the results of the Aquarius Project, an ultrahigh-resolution particle simulation of MW-sized galactic halos~\citep{2008MNRAS.391.1685S}.

\begin{sloppypar}
In order to estimate the two unknown Einasto profile parameters $\rho_{-2}$ and $r_{-2}$, we employ two constraints: first, the integration of the profile along the l.o.s., as defined in Eq.~\ref{eq:jfactor_definition}, should reproduce the value of \jann\ provided by the scaling relations. The second constraint is related to the tidal radius $r_t$ of Tri~II, or the radius at which the DM density of Tri~II is equal to the local DM density of the MW halo $\rho^{\text{MW}}_{\text{DM}}$. This criterion can be used as a benchmark to describe the extent of MW subhalos~\citep[see e.g.][]{2015MNRAS.453..849B}. In particular, we adopt equation 19 from~\cite{2015MNRAS.453..849B}:  
\begin{equation}\label{eq:tidal_radius}
	\rho^{\text{TriII}}_{\text{DM}}(r_t) = \rho^{\text{MW}}_{\text{DM}}(d_{\text{GC}} - r_t) \, ,
\end{equation}
where ${d_{\text{GC}} = 36~\si{kpc}}$ is the galactocentric distance of Tri~II~\citep{2015ApJ...814L...7K} and $\rho^{\text{TriII}}_{\text{DM}}(r_t)$ is the Einasto density of Tri~II evaluated at its tidal radius.  We adopt a value of ${r_t = 5^{+1}_{-3}~\si{kpc}}$ for Tri~II following~\cite{2019MNRAS.482.3480P}, as this range represents the median $r_t$ of systems with unresolved kinematics in their sample. The density of the MW DM halo at the tidal radius of Tri~II as determined by the Navarro-Frenk-White model of~\cite{2013JCAP...07..016N} is ${\log_{10}[\rho^{\text{MW}}_{\text{DM}}(d_{\text{GC}} - r_t) / \si{M_{\odot}}~\si{kpc}^{-3}] = 5.93^{+0.23}_{-0.31}}$. We use Eqs.~\ref{eq:jfactor_definition} and~\ref{eq:tidal_radius} to calculate $\rho_{-2}$ and $r_{-2}$, while fixing \jann, $r_t$, and ${\rho^{\text{MW}}_{\text{DM}}}$. The integrals of Eq.~\ref{eq:jfactor_definition} are computed using the \texttt{CLUMPY} code~\citep{2012CoPhC.183..656C,2016CoPhC.200..336B,2019CoPhC.235..336H}. Solving this system numerically yields final values for the Einasto profile parameters of ${\log_{10} [\rho_{-2} / \si{M_{\odot}}~\si{kpc}^{-3}] = 8.53^{+1.50}_{-0.09}}$ and ${r_{-2} = 0.0714^{+0.0386}_{-0.0662}~\si{kpc}}$. For this result, more than 86\% of the total integrated \emph{J}-factor signal is contained within the region $\theta^{2} = 0.02~\si{deg}^{2}$ about the center of the source.

\end{sloppypar}


\subsection{Likelihood analysis} 
\label{sub:dark_matter_analysis}

We used the full likelihood method, developed by~\cite{2012JCAP...10..032A} in conjunction with the binned likelihood analysis scheme presented in~\cite{2018JCAP...03..009A}. The DM signal was modeled with the PYTHIA simulation package version 8.135 with electroweak corrections~\citep{2011JCAP...03..051C}. The likelihood $\mathcal{L}$ is the product of two likelihood functions, one for each telescope pointing direction $i=1,2$. The binned likelihood for each pointing is written as:

\begin{alignat}{2}
    \label{eq:binned_likelihood}
	\mathcal{L}_{i} = & \mathcal{L}_{i}\bigl(\langle \sigma v \rangle; \{b_{ij}\}_{j=1,\dots,N_{\text{bins}}}, J_{\text{ann}},\, \tau_i\, | (N_{\text{ON},ij}, N_{\text{OFF},ij})_{j=1,\dots,N_{\text{bins}}}\bigr) \nonumber \\
	= & \prod_{j=1}^{N_{\text{bins}}} \biggl[ \frac{(g_{ij}(\langle \sigma v \rangle) +b_{ij})^{N_{\text{ON},ij}}}{N_{\text{ON},ij}!} e^{-(g_{ij}(\langle \sigma v \rangle)+b_{ij})}  \nonumber \\
	& \times \frac{(\tau_i b_{ij})^{N_{\text{OFF},ij}}}{N_{\text{OFF},ij}!} e^{-(\tau_ib_{ij})} \biggr] \\
    & \times  \mathcal{T}(\tau_i|\tau_{\text{obs},i},\sigma_{\tau,i}) \times \mathcal{J}(J_\text{ann}|J_{\text{obs}},\sigma_{\log_{10}J_{\text{obs}}}) \nonumber
\end{alignat}
where
\begin{itemize}
	\item ${N_{\text{bins}}}$ is the number of considered bins of the estimated energy;
	\item ${g_{ij}}$ and ${b_{ij}}$ are the estimated number of signal and background events respectively;
	\item ${N_{\text{ON},ij}}$ and ${N_{\text{OFF},ij}}$ are the number of observed events in the $j$-th~ON and OFF bins respectively, where OFF events are obtained from the analogous ON region at the complementary pointing;
	\item ${\mathcal{J}}$ is the likelihood for the annihilation J-factor $J_{ann}$, given measured $\log_{10}J_{obs}$ and its uncertainty $\sigma_{\log_{10}J_{\text{obs}}}$, as defined by the equation
	\begin{equation}\label{J_likelihood}
		\mathcal{J}(J_{\text{ann}}|J_{\text{obs}},\sigma_{\log_{10}J_{obs}})=\frac{1}{\ln(10)J_{\text{obs}}\sqrt{2\pi}\sigma_{\log_{10}J_{\text{obs}}}} \times e^{-\bigl(\log_{10}(J_{\text{ann}})-\log_{10}(J_{\text{obs}})\bigr)^{2}/2\sigma_{\log_{10}J_{\text{obs}}}^{2}}\ ;
	\end{equation}
	\item ${\mathcal{T}}$ is the likelihood for ${\tau_i}$ (the ON/OFF) acceptance ratio, parameterized by a Gaussian function with mean ${\tau_{\text{obs},i}}$ and variance ${\sigma^{2}_{\tau,i}}$, which includes statistical and systematic uncertainties, added in quadrature assuming Poisson statistics. We consider a systematic uncertainty for the parameter ${\tau_{\text{sys},i}}=0.015{\tau_{\text{obs},i}}$, a value that has been established in~\cite{2014JCAP...02..008A}.	
\end{itemize}
The parameters \jann, ${\tau_i}$, and ${b_{ij}}$ are nuisance parameters. The variable ${g_{ij}}$ depends on the free parameter ${\langle \sigma v \rangle}$ as such:
\begin{equation}
	{g_{ij}}\left(\langle \sigma v \rangle\right) = T_{\text{obs},i} \int_{E'_{\text{min},j}}^{{E'_{\text{max},j}}} dE' \int_{0}^{\infty} dE \frac{d\Phi_{\gamma}\left(\langle \sigma v \rangle\right)}{dE} A_{\text{eff}}(E) G(E'|E),
\end{equation}
where ${T_{\text{obs},i}}$ is the total observation time, and $E$ and $E'$ are the true and estimated gamma-ray energy, respectively. ${E'_{\text{min},j}}$ and ${E'_{\text{max},j}}$ are the minimum and maximum energies of the j-th energy bin, respectively. $A_{\text{eff}}$ is the effective collection area, and $G$ represents the probability density function of the energy estimator, both computed from a MC simulated gamma-ray data set following the spatial distribution expected from signals induced by the annihilation of DM in Tri~II computed in Section~\ref{sub:determination_of_jann} (through a MAGIC dedicated procedure described in~\cite{2018JCAP...03..009A}). 

\subsection{Results on dark matter annihilation models} 
\label{sec:results_on_dark_matter_annihilation_models}

We performed a search for annihilating DM in the candidate dSph Tri~II using 62.4 h of good-quality data, assuming DM particles with masses between 200~GeV and 200~TeV, and a 100\% branching ratio into four different channels: \HepProcess{\Pbottom\APbottom}, \HepProcess{\APmuon\Pmuon}, \HepProcess{\APtauon\Ptauon}, and \HepProcess{\PWp\PWm}.
\begin{figure}[!tb]
	\centering
 	\begin{subfigure}{.48\textwidth}
		\centering
		\includegraphics[width=\linewidth]{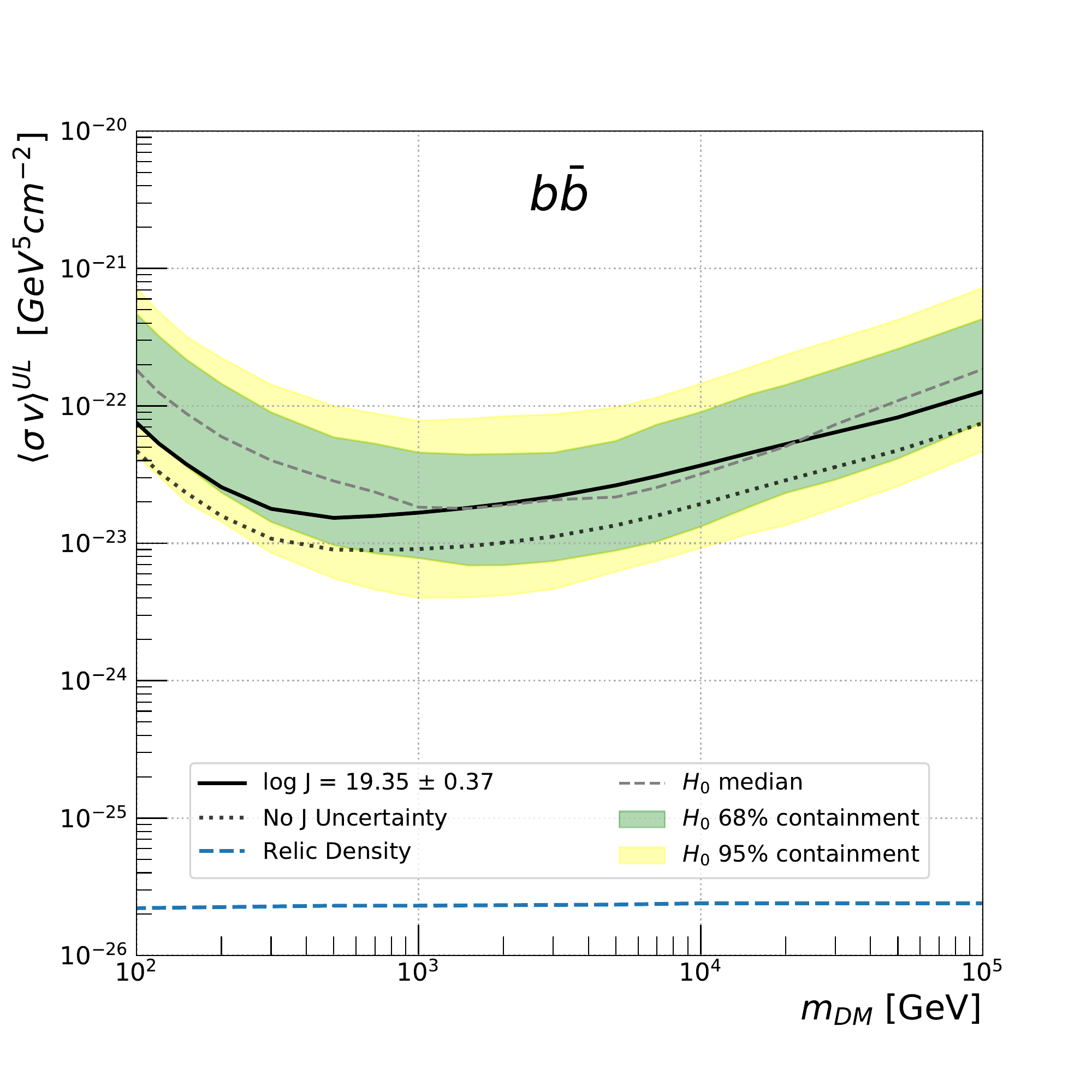}
	\end{subfigure}\hfill
 	\begin{subfigure}{.48\textwidth}
		\centering
		\includegraphics[width=\linewidth]{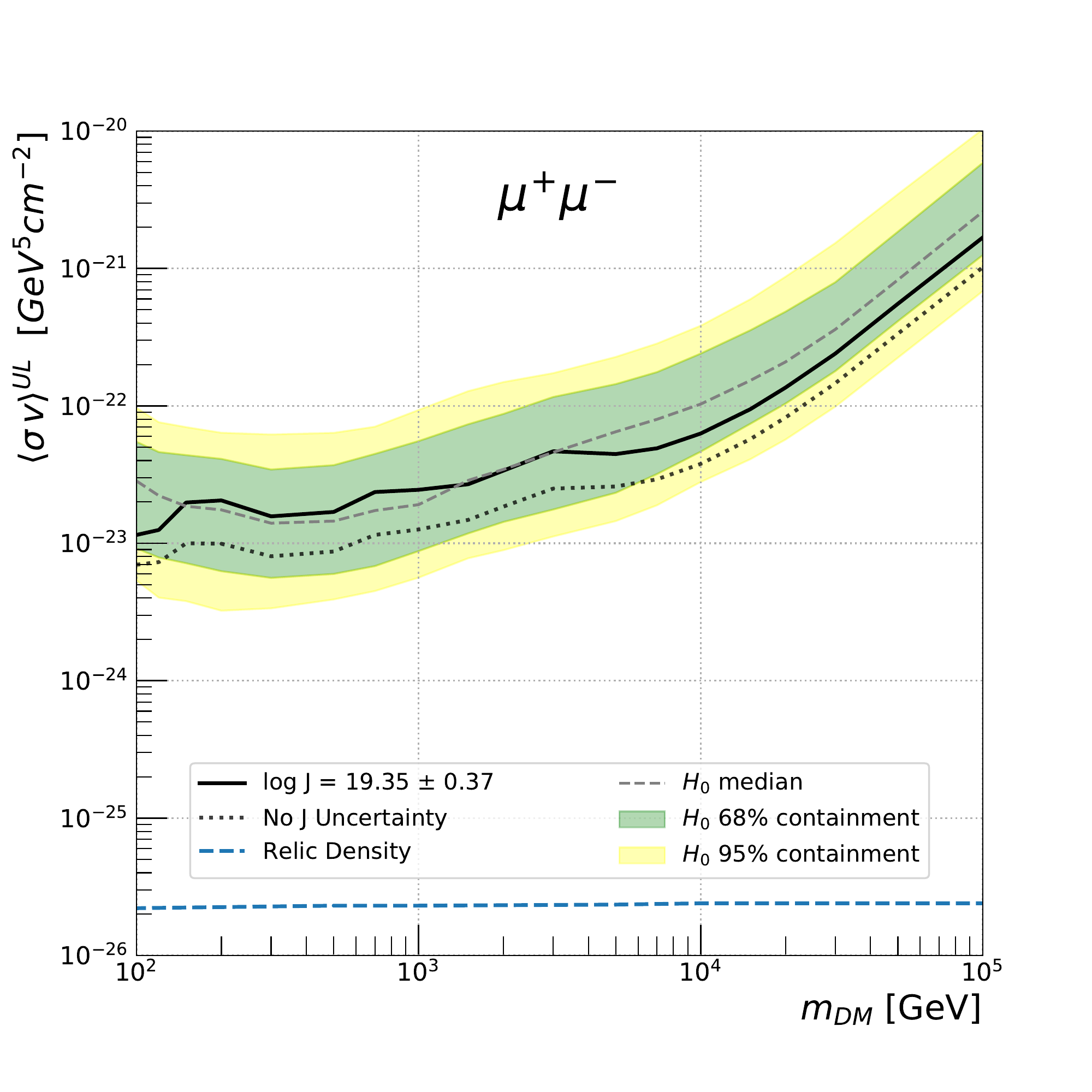}
	\end{subfigure}\hfill
 	\begin{subfigure}{.48\textwidth}
		\centering
		\includegraphics[width=\linewidth]{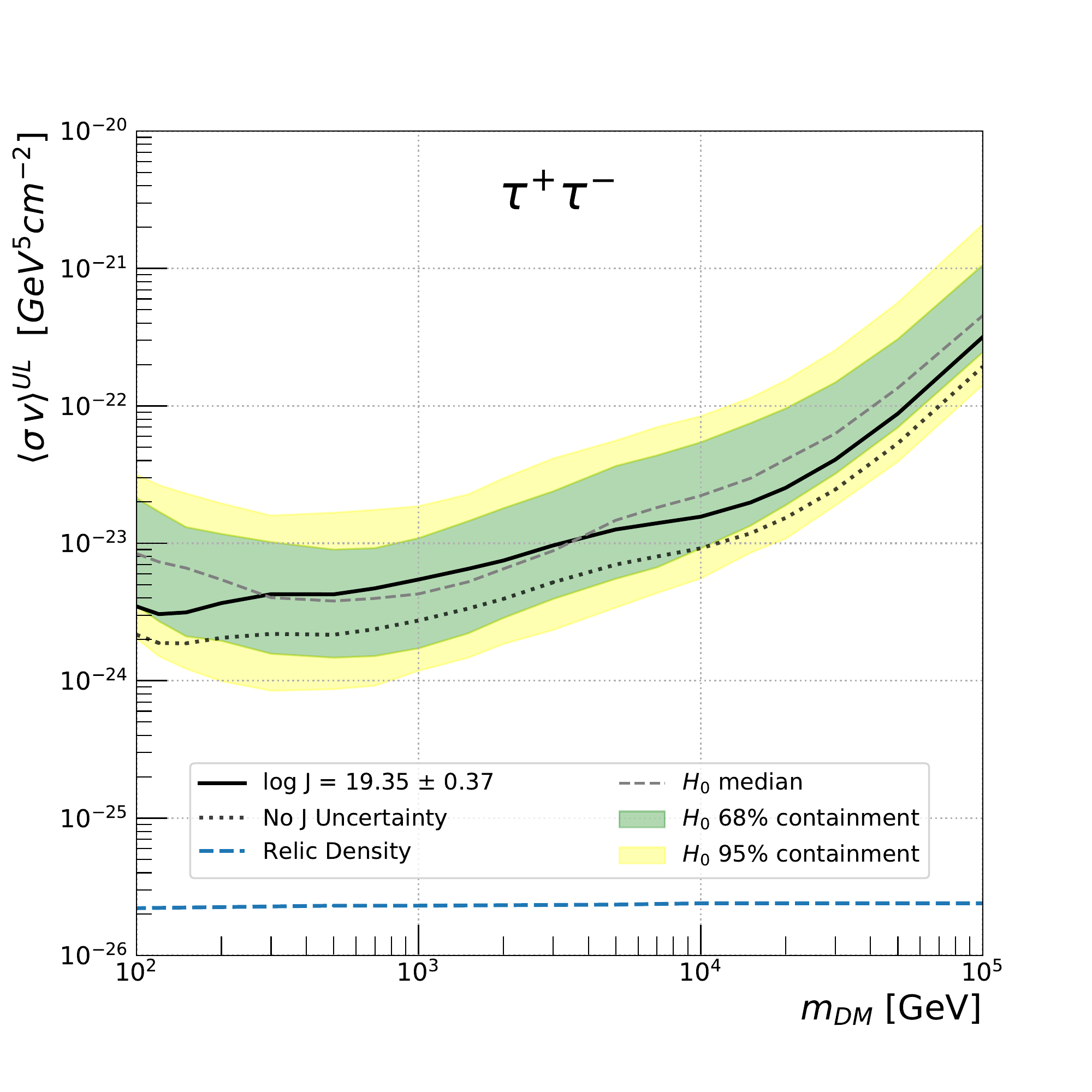}
	\end{subfigure}\hfill
 	\begin{subfigure}{.48\textwidth}
		\centering
		\includegraphics[width=\linewidth]{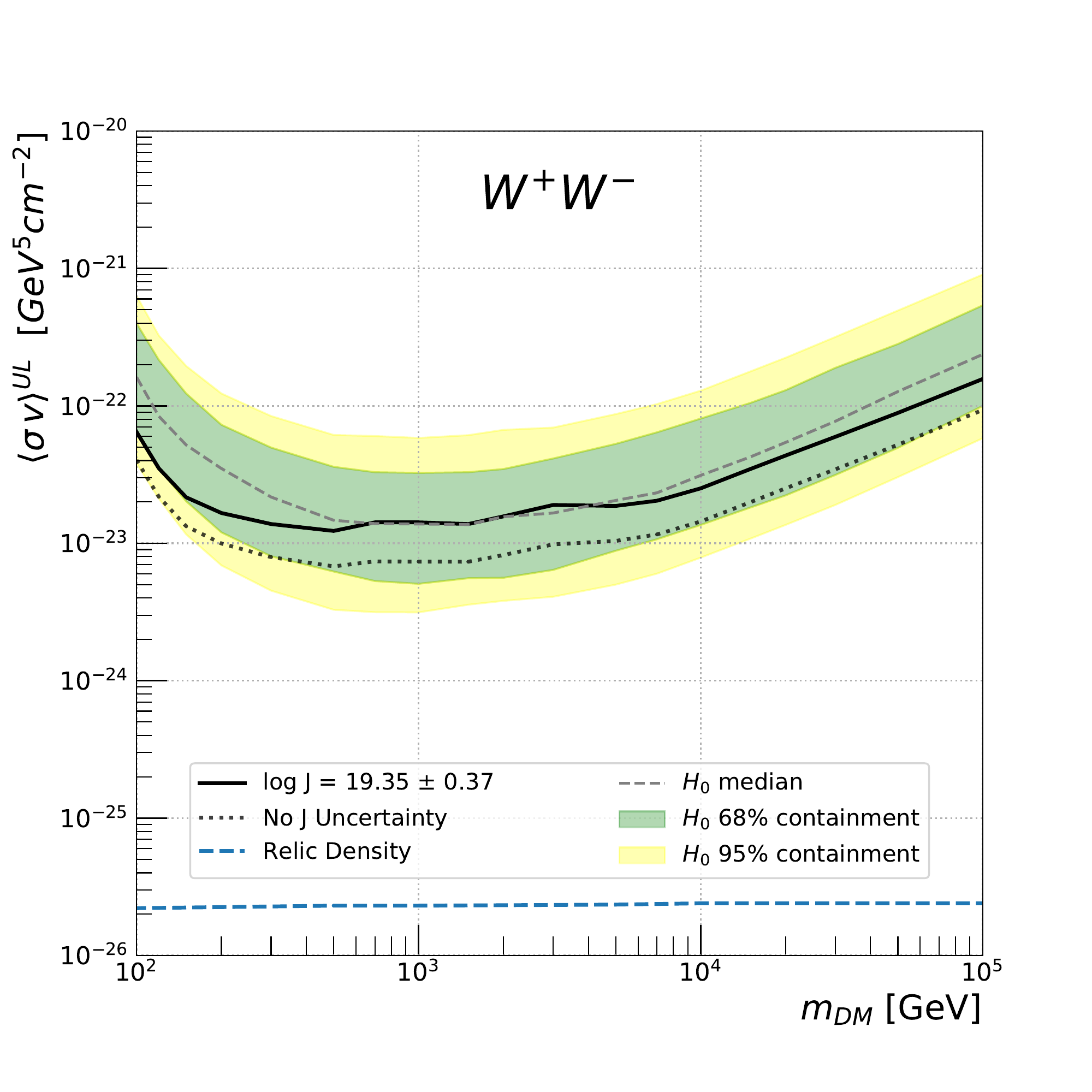}
	\end{subfigure}\hfill
	\caption{95\% CL ULs on the DM annihilation cross section ${\langle \sigma v \rangle}$ as a function of the dark matter mass $m_{DM}$ for four different channels: \HepProcess{\Pbottom\APbottom} (\emph{upper left}), \HepProcess{\APmuon\Pmuon} (\emph{upper right}), \HepProcess{\APtauon\Ptauon} (\emph{bottom left}), and \HepProcess{\PWp\PWm} (\emph{bottom right}). The extended \emph{J}-factor profile calculated in Section~\ref{sub:determination_of_jann}, with a total value of ${\log[J_{\text{ann}}({0.5^{\circ}})/~\si{GeV^{2}}\si{cm^{-5}}] = 19.35 \pm 0.37}$, is used in the likelihood to calculate the limits. The observed (solid line) and the expected (dashed line) limits are shown together with the $1\sigma$ (green) and $2\sigma$ (yellow) bands. The dotted line shows the limits calculated with the same \emph{J}-factor but assuming no uncertainty. The thermal relic cross section (blue dash-dotted) is also shown~\citep{2012PhRvD..86b3506S}.}
	\label{fig:annihilation_limits}
\end{figure}
We present the 95\% CL ULs on the thermally-averaged DM annihilation cross section ${\langle \sigma_{\text{ann}} v \rangle}$ for each annihilation channel with a binned likelihood analysis,\footnote{We construct and minimize our likelihood making use of \texttt{https://github.com/javierrico/gLike}} with 10 logarithmically-spaced bins ranging from 80 GeV to 10 TeV.\footnote{Empty bins were merged with neighboring ones.} We considered a \emph{J}-factor of ${\log_{10}[\jann(0.5^{\circ}) / \si{GeV^2}\si{cm}^{-5}] = 19.35 \pm 0.37}$ as obtained in Section~\ref{sub:determination_of_jann}. Results for leptonic and hadronic channels are shown in Figure~\ref{fig:annihilation_limits}. 

The expected limits from the null hypothesis of ${\langle \sigma v \rangle} = 0$ (dashed lines) and its 68\% and 95\% containment bands (in green and yellow, respectively) are also shown. These bands are calculated with 300 MC simulations in which both ON and OFF regions are generated from pure background probability density functions while assuming a similar exposure for the real data. In this case, $\tau_{i}$ is taken as a nuisance parameter in the likelihood function. Our limits are consistent with the null hypothesis for all considered DM models. The most constraining limits are ${\langle \sigma v \rangle^{\text{UL}} = 1.53 \times 10^{-23}}$~\si{cm^{3} s^{-1}} in the \HepProcess{\Pbottom\APbottom} channel for $m_{DM} = 500$~\si{GeV} and ${\langle \sigma v \rangle^{\text{UL}} = 3.05 \times 10^{-24}}$~\si{cm^{3} s^{-1}} in the \HepProcess{\APtauon\Ptauon} channel for $m_{DM} = 120$~\si{GeV}. The limits assuming a fixed \jann\ with no uncertainly, calculated by ignoring the \emph{J}-factor term in Eq.~\ref{eq:binned_likelihood}, are also shown (dotted lines).

The Tri~II results for the \HepProcess{\APtauon\Ptauon} channel are compatible with previous MAGIC dSph studies, including those derived from 94.8~\si{h} and 157.9~\si{h} of Ursa~Major~II~\cite{2018JCAP...03..009A} and Segue~I~\cite{2014JCAP...02..008A} observations, respectively. Due to the difference in the analysis method developed and adopted after the Segue~I results, in both the treatment of the \emph{J}-factor as a nuisance parameter and in the background modeling methods, a straightforward comparison is not easily achievable. However, a comparison is made possible by looking at the MAGIC-only Segue~1 results shown in a later combined analysis between Fermi and MAGIC~\cite{2016JCAP...02..039M} using the same data set. The differences with respect to this later work can be explained taking into account the difference in exposures and \emph{J}-factor between both analyses.
Results from different dSphs can be combined into a unique limit using a specific recipe described in~\cite{2012JCAP...10..032A}. Some preliminary results were shown in~\cite{2019ICRC...36..539O} and will be published in a future manuscript.



\section{Summary and Conclusions} 
\label{sec:summary_and_conclusions}

\begin{sloppypar}
An observation campaign of Triangulum~II was carried out with MAGIC starting in August 2016 as a reaction to the announcement of a potential large quantity of DM content within the source~\citep{2015ApJ...814L...7K}. A total of 62.4~\si{h} of high-quality data were collected, and no significant gamma-ray signal was detected in both the standard point-like and extended-profile DM analyses. Integral and differential upper limits on the gamma-ray flux were obtained for the first time for the region of Tri~II. These limits can be used to guide future searches in the source region, and act as a baseline for further studies with next generation IACTs~\citep{2019scta.book.....C}.

A dedicated search for an extended DM source was carried out in the region. Due to the limited available stellar kinematic data for Tri~II, a method based on scaling relations was used to infer the DM content of the object, resulting in a \emph{J}-factor of ${\log_{10}[\jann(0.5^{\circ}) / \si{GeV^2}\si{cm}^{-5}] = 19.35 \pm 0.37}$. Based on knowledge provided by numerical simulations, we derived a DM profile for Tri~II which we subsequently used in a rigorous DM binned likelihood analysis. The obtained 95\% confidence level upper limits on the DM annihilation cross section are compatible with the null hypothesis 68\% containment bands. In addition, the use of the maximum likelihood method would allow for the combination of our data with those available for additional dSphs, even if observed with other instruments. These combined analyses will help to improve the current sensitivity of DM searches towards the thermal relic limit in the $>$1~\si{TeV} mass range~\citep[e.g.][]{2016JCAP...02..039M}. 

With the sparse amount of empirical knowledge surrounding the DM content of Tri~II presently available in the literature, we believe an approach that does not rely on spectroscopic data is currently the most reliable option. Ideally, in the near future, a more precise determination of the DM annihilation limits derived from observations of Tri~II can be obtained using the same MAGIC data set as soon as the behavior of its stellar kinematics is better understood.
\end{sloppypar}

\appendix
\begin{landscape}
\section{Gamma-ray flux upper limits} 
\label{sec:gamma_ray_flux_upper_limits}

Tables~\ref{tab:Differential_flux_upper_lims} and~\ref{tab:Integral_flux_upper_lims} show the differential and integral upper limits, respectively, on the gamma-ray flux from Tri~II assuming a point-like source and power-law spectra with various photon indices $\Gamma$. For more details see Section~\ref{sub:flux_upper_limits}. In both tables, $N_{\textrm{ON}}$ and $N_{\textrm{OFF}}$ are the total number of observed events in the ON and OFF regions, respectively. The normalization factor between the ON and OFF regions is represented by $\tau$. Finally, $N^{\textrm{UL}}_{\textrm{ex}}$ is the upper limit on the number of excess events calculated for a 95\% CL with the conventional method~\citep{2005NIMPA.551..493R}, assuming a systematic uncertainty on the overall detection efficiency of 30\% (a standard value for MAGIC analyses).

	\setcaptiontype{table}
	\centering
	\vspace*{\fill} 
	\resizebox{1.25\textwidth}{!}{%
	\begin{tabular}{cccccccccccc}
		\toprule
        & & & & & \multicolumn{7}{c}{$d\Phi/dE^{\textrm{UL}}$ [ \si{TeV^{-1} cm^{-2} s^{-1}} ]}  \\ 
		\cmidrule{6-12}
		$\Delta E$ [GeV] & $N_{\textrm{ON}}$ & $N_{\textrm{OFF}}$ & $\tau$ & $N^{\textrm{UL}}_{\textrm{ex}}$ & $\Gamma = -1.0$ & $\Gamma = -1.2$ & $\Gamma = -1.5$ & $\Gamma = -1.8$ & $\Gamma = -2$ & $\Gamma = -2.2$ & $\Gamma = -2.4$  \\                                                     
		\toprule
		100 -- 316.2     & 6854  & 6844 & 1.002 & 309.3 & $4.1\times10^{-12}$ & $4.2\times10^{-12}$ & $4.4\times10^{-12}$ & $4.5\times10^{-12}$ & $4.6\times10^{-12}$ & $4.7\times10^{-12}$ & $4.8\times10^{-12}$  \\
		316.2 -- 1000    & 967   & 1003 & 1.011 & 71.8  & $4.6\times10^{-13}$ & $4.7\times10^{-13}$ & $4.8\times10^{-13}$ & $4.9\times10^{-13}$ & $5.0\times10^{-13}$ & $5.1\times10^{-13}$ & $5.1\times10^{-13}$  \\
		1000 -- 3162.2   & 113   & 120  & 1.005 & 28.4  & $1.4\times10^{-13}$ & $1.4\times10^{-13}$ & $1.4\times10^{-13}$ & $1.5\times10^{-13}$ & $1.5\times10^{-13}$ & $1.5\times10^{-13}$ & $1.5\times10^{-13}$  \\
		3162.2 -- 10000  & 7     & 9    & 1.005 & 7.4   & $3.8\times10^{-14}$ & $3.9\times10^{-14}$ & $4.0\times10^{-14}$ & $4.1\times10^{-14}$ & $4.1\times10^{-14}$ & $4.2\times10^{-14}$ & $4.2\times10^{-14}$  \\
		\bottomrule                                                 
	\end{tabular}}
	\caption{Differential gamma-ray flux upper limits of Tri~II.}
	\label{tab:Differential_flux_upper_lims}

	\bigskip\bigskip\bigskip\bigskip  

	\resizebox{1.25\textwidth}{!}{%
	\begin{tabular}{cccccccccccc} 
		\toprule
		&  &  &  &  & \multicolumn{7}{c}{$\Phi^{\textrm{UL}}$ [ \si{cm^{-2}.s^{-1}} ]}  \\ 
		\cmidrule{6-12}
		$E_{\textrm{th}}$ [GeV] & $N_{\textrm{ON}}$ & $N_{\textrm{OFF}}$ & $\tau$ & $N^{\textrm{UL}}_{\textrm{ex}}$ & $\Gamma = -1.0$ & $\Gamma = -1.2$ & $\Gamma = -1.5$ & $\Gamma = -1.8$ & $\Gamma = -2$ & $\Gamma = -2.2$ & $\Gamma = -2.4$  \\ 
		\toprule
		100    & 7942 & 7978  & 1.002  & 269.3  & $2.5\times10^{-12}$ & $2.3\times10^{-12}$ & $2.3\times10^{-12}$ & $2.6\times10^{-12}$ & $2.9\times10^{-12}$ & $3.1\times10^{-12}$ & $3.3\times10^{-12}$   \\
		177.8  & 2930 & 2918  & 1.002  & 211.3  & $1.9\times10^{-12}$ & $1.6\times10^{-12}$ & $1.6\times10^{-12}$ & $1.6\times10^{-12}$ & $1.7\times10^{-12}$ & $1.7\times10^{-12}$ & $1.8\times10^{-12}$   \\
		316.2  & 1088 & 1133  & 1.011  & 70.9   & $6.3\times10^{-13}$ & $5.2\times10^{-13}$ & $4.7\times10^{-13}$ & $4.5\times10^{-13}$ & $4.6\times10^{-13}$ & $4.7\times10^{-13}$ & $4.8\times10^{-13}$   \\
		562.3  & 380  & 402   & 1.008  & 44.5   & $4.1\times10^{-13}$ & $3.3\times10^{-13}$ & $2.8\times10^{-13}$ & $2.6\times10^{-13}$ & $2.6\times10^{-13}$ & $2.6\times10^{-13}$ & $2.6\times10^{-13}$   \\
		1000   & 121  & 130   & 1.005  & 28.6   & $2.8\times10^{-13}$ & $2.2\times10^{-13}$ & $1.9\times10^{-13}$ & $1.7\times10^{-13}$ & $1.6\times10^{-13}$ & $1.6\times10^{-13}$ & $1.6\times10^{-13}$   \\
		1778.3 & 29   & 31    & 1.005  & 16.6   & $1.9\times10^{-13}$ & $1.5\times10^{-13}$ & $1.2\times10^{-13}$ & $1.1\times10^{-13}$ & $1.0\times10^{-13}$ & $9.8\times10^{-14}$ & $9.6\times10^{-14}$   \\
		3162.3 & 8    & 9     & 1.005  & 8.9    & $1.3\times10^{-13}$ & $9.8\times10^{-14}$ & $8.0\times10^{-14}$ & $6.4\times10^{-14}$ & $6.1\times10^{-14}$ & $5.8\times10^{-14}$ & $5.6\times10^{-14}$   \\
		5623.4 & 2    & 1     & 1.000  & 6.7    & $1.7\times10^{-13}$ & $1.1\times10^{-13}$ & $8.5\times10^{-14}$ & $6.1\times10^{-14}$ & $5.6\times10^{-14}$ & $5.2\times10^{-14}$ & $4.9\times10^{-14}$   \\
		\bottomrule
	\end{tabular}}
	\caption{Integral gamma-ray flux upper limits of Tri~II for events with energies above various energy thresholds $E_{\textrm{th}}$.}
	\label{tab:Integral_flux_upper_lims}
	\vspace*{\fill} 
\end{landscape}


\section*{Acknowledgments} 
\label{sec:acknowledgments}
\begin{sloppypar}
We would like to thank Miguel Angel S\'anchez-Conde for his assistance interpreting the Aquarius Project simulations.

We would also like to thank the Instituto de Astrof\'{\i}sica de Canarias for the excellent working conditions at the Observatorio del Roque de los Muchachos in La Palma. The financial support of the German BMBF and MPG, the Italian INFN and INAF, the Swiss National Fund SNF, the ERDF under the Spanish MINECO (FPA2017-87859-P, FPA2017-85668-P, FPA2017-82729-C6-2-R, FPA2017-82729-C6-6-R, FPA2017-82729-C6-5-R, AYA2015-71042-P, AYA2016-76012-C3-1-P, ESP2017-87055-C2-2-P, FPA2017‐90566‐REDC), the Indian Department of Atomic Energy, the Japanese JSPS and MEXT, the Bulgarian Ministry of Education and Science, National RI Roadmap Project DO1-153/28.08.2018 and the Academy of Finland grant nr. 320045 is gratefully acknowledged. This work was also supported by the Spanish Centro de Excelencia ``Severo Ochoa'' SEV-2016-0588 and SEV-2015-0548, and Unidad de Excelencia ``Mar\'{\i}a de Maeztu'' MDM-2014-0369, by the Croatian Science Foundation (HrZZ) Project IP-2016-06-9782 and the University of Rijeka Project 13.12.1.3.02, by the DFG Collaborative Research Centers SFB823/C4 and SFB876/C3, the Polish National Research Centre grant UMO-2016/22/M/ST9/00382 and by the Brazilian MCTIC, CNPq and FAPERJ.
\end{sloppypar}


\bibliographystyle{elsarticle-num}
\bibliography{biblio}


\end{document}